\documentclass[showpacs,aps,prl,reprint]{revtex4-1}

\usepackage{graphicx,psfrag,epsf,epsfig}
\usepackage{dcolumn}
\usepackage{bm,bbm}
\usepackage{pstricks,pst-plot}     
\usepackage{latexsym}
\usepackage{mathrsfs,amsmath,amssymb,textcomp}
\usepackage{balance}
\usepackage{hyperref}

\begin{document}

\title{A Microscopic Perspective on Photovoltaic Reciprocity in Ultrathin Solar Cells } 

\author{Urs Aeberhard}
\email[]{u.aeberhard@fz-juelich.de}

\author{Uwe Rau}

\affiliation{IEK-5 Photovoltaik, Forschungszentrum J\"ulich, D-52425 J\"ulich, Germany}

\date{\today}

\begin{abstract}
The photovoltaic reciprocity theory relates the electroluminescence spectrum of a solar cell under applied bias to the external photovoltaic quantum efficiency of the device as measured at short circuit conditions. Its derivation is based on detailed balance relations between local absorption and emission rates in optically isotropic media with non-degenerate quasi-equilibrium carrier distributions. In many cases, the dependence of density and spatial variation of electronic and optical device states on the point of operation is modest and the reciprocity relation holds. In nanostructure-based photovoltaic devices exploiting confined modes, however, the underlying assumptions are no longer justifiable. In the case of ultrathin absorber solar cells, the modification of the electronic structure with applied bias is significant due to the large variation of the built-in field. Straightforward use of the external quantum efficiency as measured at short circuit conditions in the photovoltaic reciprocity theory thus fails to reproduce the electroluminescence spectrum at large forward bias voltage. This failure is demonstrated here by numerical simulation of both spectral quantities at normal incidence and emission for an ultrathin GaAs $p$-$i$-$n$ solar cell using an advanced quantum kinetic formalism based on non-equilibrium Green's functions of coupled photons and charge carriers. While coinciding with the semiclassical relations under the conditions of their validity, the theory provides a consistent microscopic relationship between absorption, emission and charge carrier transport in photovoltaic devices at arbitrary operating conditions and for any shape of optical and electronic density of states. 
\end{abstract}

\pacs{}

\maketitle 

Absorption and emission of light by a semiconductor material are two physical processes that are related by the fundamental laws of light-matter coupling. For a given set of electronic states connected by an optical transition, the two processes are described in terms of identical optical matrix elements and joint density of electronic states participating in the transition. In thermal equilibrium, the principle of detailed balance dictates a vanishing net transition rate. On these grounds, the absorption coefficient has been related to the radiative lifetime of charge carriers under non-equilibrium conditions \cite{roosbroeck:54}, and to the local emission spectrum at finite splitting of quasi-Fermi levels \cite{wuerfel:82}. In the first case, the absorption coefficient $\alpha$ determines the prefactor $\mathcal{B}$ of the local rate of radiative recombination $R(\mathbf{r})=\mathcal{B}(\mathbf{r})\rho_{n}(\mathbf{r})\rho_{p}(\mathbf{r})/n_{i}^{2}$ via
\begin{align}
\mathcal{B}(\mathbf{r})=&\int dE_{\gamma}~\alpha(\mathbf{r},E_{\gamma})n_{r}^2(\mathbf{r})\bar{\phi}_{\textrm{bb}}(E_{\gamma}),\label{eq:vrs}
\end{align}
 where $E_{\gamma}=\hbar\omega$ is the photon energy, $\rho_{n(p)}$ and $n_{i}$ denote the electron (hole) and intrinsic carrier densities, respectively, $n_{r}$ is the local refractive index -- approximated as independent of energy --  and $\bar{\phi}_{\textrm{bb}}\equiv 4\pi\phi_{\textrm{bb}}$ is the angular integration of the vacuum black-body radiation flux $\phi_{\textrm{bb}}(E_{\gamma})=E_{\gamma}^2/(4\pi^3\hbar^3c_{0}^2)f_{\textrm{BE}}(E_{\gamma})$, with $f_{\textrm{BE}}(E_{\gamma})=\left\{\exp[\beta E_{\gamma}]-1\right\}^{-1}$ ($\beta\equiv \{k_{\textrm{B}}T\}^{-1}$) the Bose-Einstein distribution function at emitter temperature $T$. In the second case, a generalization of the Planck emission law is given by the following rate (per unit volume):
 \begin{align}
\mathcal{R}^{\mathrm{GP}}(\mathbf{r},E_{\gamma},\Delta\mu)=\alpha(\mathbf{r},E_{\gamma})\bar{D}_{0}^{\gamma}(\mathbf{r},E_{\gamma})\frac{c_{0}}{n_{r}(\mathbf{r})}f_{\textrm{BE}}(E_{\gamma}-\Delta\mu),\label{eq:genplanck}
\end{align}
where $\bar{D}_{0}^{\gamma}$ is the angle-integrated density of photon states and $\Delta\mu$ denotes the quasi-Fermi level splitting (QFLS).  In  Ref.~\onlinecite{wuerfel:82}, this result was derived under the assumption of  unrestricted optical transitions (no momentum selection rule), of quasi-equilibrium occupation described by Fermi statistics with distinct and constant (bulk) quasi-Fermi levels for electrons and holes, and of an optically homogeneous medium exhibiting a photon density of states for free field modes, $\bar{D}_{0}^{\gamma}(\mathbf{r},E_{\gamma})=(E_{\gamma}^2n_{r}^3)/(\pi^2\hbar^3 c_{0}^3)$. The two cases coincide if Boltzmann statistics are used for the carrier densities in Ref.~\onlinecite{roosbroeck:54} and substituted for the Bose factor in \eqref{eq:genplanck}, which is often found in literature, though strictly valid only at high temperature and far from degeneracy. Expression \eqref{eq:genplanck} constitutes one of the main ingredients of the photovoltaic reciprocity theory as formulated in Ref.~\onlinecite{rau:07}, which relates the luminescent emission of a solar cell under applied bias voltage to the external quantum efficiency $Q_{\mathrm{PV}}$ of the same device under illumination, and which has found widespread application in luminescence-based characterization of photovoltaic materials, cells and modules \cite{kirchartz:07_jap,kirchartz:08_apl,kirchartz:09_pip,hoyer:10}. For the case of an applied electrical bias $V$ and a $Q_{\textrm{PV}}$ that is determined at normal incidence and short circuit conditions ($V=0$), this reciprocity relation reads \cite{rau:07}
\begin{align}
\phi_{\mathrm{em}}^{\perp}(E_{\gamma},V)=&\,Q_{\textrm{PV}}^{\perp}(E_{\gamma})\phi_{bb}(E_{\gamma})\left[\exp\left(\frac{qV}{k_{B}T}\right)-1\right].\label{eq:recipro_norm}
\end{align}

\begin{figure}[t]
	\begin{center}
		\includegraphics[width=0.45\textwidth]{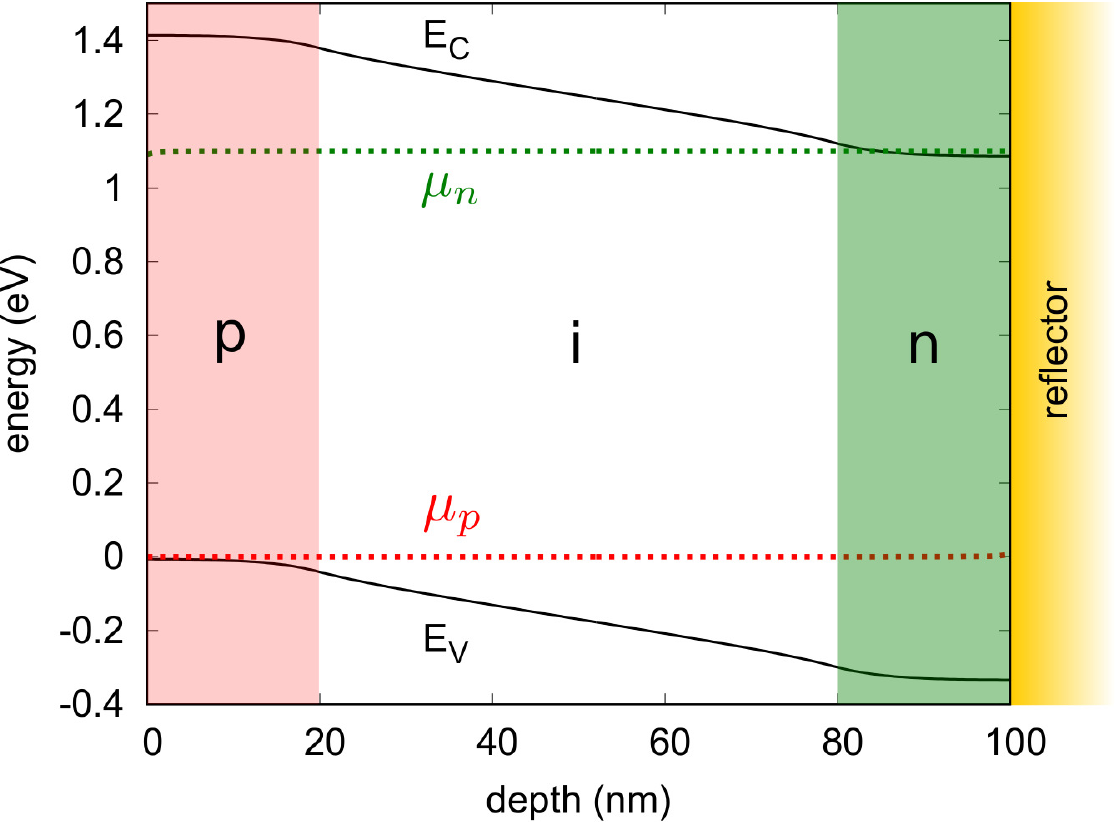}
		\caption{(color online) Band profile - conduction band edge E$_{\textrm{C}}$ and valence band edge E$_{\textrm{V}}$ - and quasi-Fermi levels for electrons ($\mu_{\textrm{n}}$) and holes ($\mu_{\textrm{p}}$) in a 100 nm thin GaAs $p$-$i$-$n$ - diode at applied bias voltage  of $V=1.1$ V. Both quantities are obtained from the solution of the full NEGF-Poisson problem. \label{fig:bandprof}}
	\end{center}
\end{figure}

The limits for the applicability of the photovoltaic reciprocity relation as encoded in Expr.~\eqref{eq:recipro_norm} have been investigated for different situations where some of the basic assumptions -  such as, the superposition principle or the Donolato theorem \cite{donolato:85} - do not hold \cite{kirchartz:08_pss,kirchartz:08_prb,wang:13,kirchartz:16_prappl}. For instance, in Ref.~\onlinecite{wang:13}, numerical device simulations were used to show the failure of the reciprocity relation in the case of non-identical band diagrams in the dark and under illumination. However, these numerical experiments are limited by the validity of the underlying description of the photovoltaic device operation, which has thus far been restricted to semiclassical bulk physics. On the other hand, there is a range of photovoltaic device architectures under active development where this conventional bulk picture is no longer appropriate as it fails to reflect the dependence of the microscopic electronic structure on the operating point of the device, such as nanostructure-based and ultra-thin solar cells \cite{ae:jpv_16,ae:apl_16}.  In this Letter, we therefore assess the validity of Eqs. \eqref{eq:genplanck} and \eqref{eq:recipro_norm} if applied to ultra-thin solar cells by comparison with the quantum-kinetic picture of radiative charge carrier generation, transport and recombination as formulated within the non-equilibrium Green's function (NEGF) formalism \cite{ae:prb_08,ae:jcel_11,ae:oqel_14}. To this end, the coupled NEGF problems of interacting electrons and photons are solved - for the first time - for a realistic solar cell device exhibiting a complex potential profile, such as the ultra-thin GaAs $p$-$i$-$n$ structure displayed in Fig.~\ref{fig:bandprof}.

In the case of a planar absorber, the \emph{local} NEGF expressions for the spectral radiative rates (per unit volume, $z$: depth) read \cite{ae:jpe_14,ae:prb89_14} 
\begin{align}
\mathcal{R}_{\mathrm{abs/em}}(z,E_{\gamma})=&\sum_{\mu}\int \frac{d^{2}\mathbf{q}_{\parallel}}{(2\pi)^2}r_{\mathrm{abs/em}}^{\mu}(\mathbf{q}_{\parallel},z,E_{\gamma}),\label{eq:locspecrate}
\end{align}
with the modal components for net absorption $r_{\mathrm{abs,net}}\equiv r_{\mathrm{abs}}-r_{\mathrm{em,stim}}$, accounting for stimulated emission, and spontaneous emission $r_{\mathrm{em,spont}}\equiv r_{\mathrm{em}}-r_{\mathrm{em,stim}}$ given by
\begin{align}
r_{\mathrm{abs,net}}^{\mu}(\mathbf{q}_{\parallel},z,E_{\gamma})
=&~\sum_{\nu}\int dz'~\Big[\mathcal{D}^{<}_{\mu\nu}(\mathbf{q}_{\parallel},z,z',E_{\gamma})\nonumber\\&\times\hat{\Pi}_{\nu\mu}(\mathbf{q}_{\parallel},z',z,E_{\gamma})\Big]/(2\pi\hbar),\label{eq:locspecabsmodrate}\\
r_{\mathrm{em,spont}}^{\mu}(\mathbf{q}_{\parallel},z,E_{\gamma})
=&~\sum_{\nu}\int dz'~\hat{\mathcal{D}}_{\mu\nu}(\mathbf{q}_{\parallel},z,z',E_{\gamma})\nonumber\\&\times\Pi_{\nu\mu}^{<}(\mathbf{q}_{\parallel},z',z,E_{\gamma})\Big]/(2\pi\hbar).\label{eq:locspecemmodrate}
\end{align}
In the above equations, $\mathbf{q}_{\parallel}$ is the transverse -- i.e., in-plane -- component of the photon wave vector,  $\boldsymbol{\mathcal{D}}$ denotes the transverse photon Green's function and $\mathbf{\Pi}$ is the transverse photon self-energy tensor related to the polarization of the electronic system \footnote{See Supplemental Material for relation to charge carrier GF and transport properties.}. These expressions provide a general microscopic relation between local absorption and emission rates in terms of occupied photon and electron-hole pair states ($\mathcal{D}^{<},\Pi^{<}$) and the corresponding final state spectral functions ($\hat {\Pi}\equiv \Pi^{>}-\Pi^{<},\hat{\mathcal{D}}\equiv  \mathcal{D}^{>}-\mathcal{D}^{<}$). The expressions are valid for arbitrary shape of the density of states and nonequilibrium occupation corresponding to the actual operating point of the structure under optical or electronic excitation.

For the comparison with the semiclassical quasi-equilibrium result \eqref{eq:genplanck}, several approximations corresponding to the restricted regime of validity of the latter are applied to  \eqref{eq:locspecabsmodrate} and \eqref{eq:locspecemmodrate}. In the case of the absorption, coupling to a coherent radiation field permits to replace the photon GF by the electromagnetic vector potential $\mathbf{A}$ of the incident radiation field,
\begin{align}
r_{\mathrm{abs,net}}^{\mu}(\mathbf{q}_{\parallel},z,E_{\gamma})=&\frac{i}{\hbar\mu_{0}}
A_{\mu}(\mathbf{q}_{\parallel},z,E_{\gamma})\int dz'\Big[
A^{*}_{\mu}(\mathbf{q}_{\parallel},z',E_{\gamma})\nonumber\\&\times
\hat{\Pi}_{\mu\mu}(\mathbf{q}_{\parallel},z',z,E_{\gamma})\big]\label{eq:rate_coh}\\
\equiv&~\Phi_{\mu}(\mathbf{q}_{\parallel},z,E_{\gamma})	
\alpha_{\mu}(\mathbf{q}_{\parallel},z,E_{\gamma}).\label{eq:genrate_def}
\end{align} 
Equation \eqref{eq:genrate_def} formally defines the local and modal absorption coefficient $\alpha$ via the local generation rate and the local value of the photon flux $\Phi$. For slow variation of the transverse electromagnetic field in the absorber, the absorption coefficient can be expressed solely in terms of the electronic properties of the absorber using the modal form  \cite{ae:prb89_14}
\begin{align}
\alpha_{\mu}(\mathbf{q}_{\parallel},z,E_{\gamma})\approx&\frac{\hbar c_{0}}{2
n_{r}E_{\gamma}}\int dz'~i\hat{\Pi}_{\mu\mu}(\mathbf{q}_{\parallel},z',z,E_{\gamma}).
\label{eq:locabscoef}
\end{align}
To account for the isotropy of the media assumed in the derivation of \eqref{eq:genplanck} and \eqref{eq:recipro_norm}, the average local absorption coefficient for an isotropic medium is considered via 
\begin{align}
\bar{\alpha}(z,E_{\gamma})\approx&\frac{\hbar c_{0}}{2
n_{r}E_{\gamma}}\int dz' i\hat{\bar{\Pi}}(\mathbf{0},z',z,E_{\gamma}), \label{eq:abscoef_av}
\end{align}
where $\bar{\Pi}=\sum_{\mu}\Pi_{\mu\mu}/3$. Use of the zero photon momentum component is justified by the weak $\mathbf{q}_{\parallel}$-dependence of the polarization function in the relevant range of photon wave vectors, since the latter are small as compared to the charge carrier quasi-momenta.  

\begin{figure}[t]
	\begin{center}	
		\includegraphics[width=0.50\textwidth]{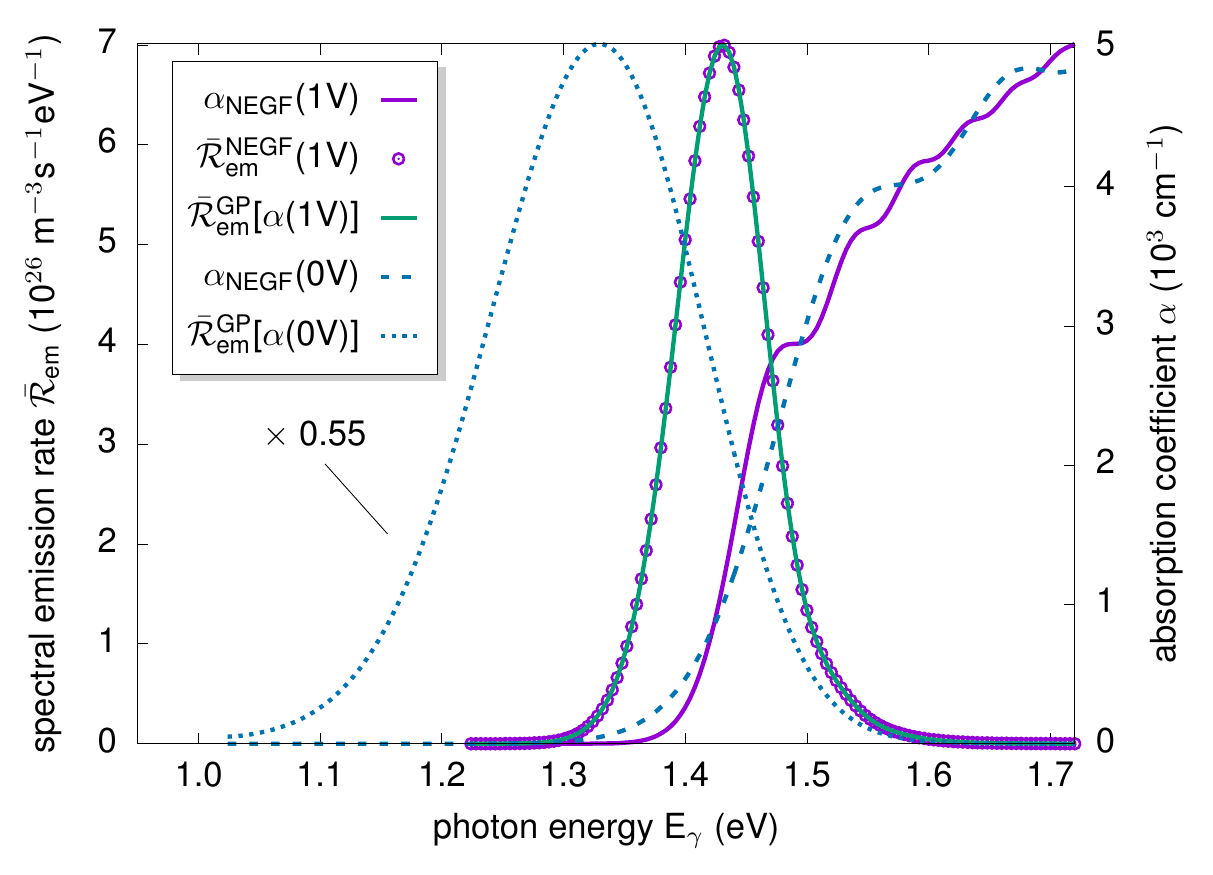}
		\caption{(color online) Local absorption coefficient $\alpha$ and spectral volume emission rate in the center of the intrinsic region of a 100\,nm thick GaAs $p$-$i$-$n$ solar cell. Dark lines represent the NEGF absorption coefficients at $V=0$\,V (dashed) and $V=1$\,V (solid), respectively. The corresponding generalized Planck (GP) spectra for QFLS of $\Delta\mu=1$\,V are given by light solid and  dotted lines, respectively. The isotropic NEGF result $\bar{\mathcal{R}}_{\textrm{em}}^{\textrm{NEGF}}$ for the emission at $V=1$\,V (open symbols) is in excellent agreement with the GP spectrum $\bar{\mathcal{R}}_{\textrm{em}}^{\textrm{GP}}$ obtained from the isotropic NEGF absorption coefficient at this bias voltage, but strongly broadened and red-shifted as compared to the GP spectrum obtained from the absorption coefficient at short circuit conditions ($V=0$\,V, dotted line).\label{fig:locabsem}}
	\end{center}
\end{figure}

In analogy to the treatment of the absorption coefficient in \eqref{eq:abscoef_av}, the average emission for an isotropic medium is found as
\begin{align}
\bar{\mathcal{R}}_{\mathrm{em}}(z,E_{\gamma})\approx&\frac{
n_{r} E_{\gamma}}{2\pi^2\hbar^2c_{0}}\int dz'~i\bar{\Pi}^{<}(\mathbf{0},z',z,E_{\gamma}),\label{eq:specemrate_av}
\end{align}
where for the last step, the momentum integration of the bare photon Green's function was used \cite{ae:jpe_14},
\begin{align}
	\int \frac{d^{2}\mathbf{q}_{\parallel}}{(2\pi)^2}\hat{\mathcal{D}}_{\mu\nu,0}(\mathbf{q}_{\parallel},z,z',E_{\gamma})=-\frac{i
	n_{r}E_{\gamma}}{3\pi\hbar c_{0}}\delta_{\mu\nu}. 
\end{align}
 At global quasi-equilibrium conditions - i.e., constant QFLS - the Kubo-Martin-Schwinger relation \cite{pereira:96} $\mathbf{\Pi}^{<}(E_{\gamma})=e^{-\beta(E_{\gamma}-\Delta\mu})\mathbf{\Pi}^{>}(E_{\gamma})$  yields $\mathbf{\Pi}^{<}(E_{\gamma})=\hat{\mathbf{\Pi}}(E_{\gamma})f_{BE}(E_{\gamma}-\Delta\mu)$, which, inserted in \eqref{eq:specemrate_av}, reproduces Expr.~\eqref{eq:genplanck}. At this point, however, it needs to be emphasized that the two polarization function components entering expressions \eqref{eq:abscoef_av} and \eqref{eq:specemrate_av} are computed on the basis of the same electronic GF, which in turn may depend strongly on the point of operation. As shown in Fig.~\ref{fig:locabsem} displaying the absorption coefficient and spectral emission rate in the center of the intrinsic region of a 100\,nm GaAs $p$-$i$-$n$ photodiode, this is indeed the case for ultrathin absorber solar cells: due to the strong impact of the built-in field on the local absorption coefficient \cite{ae:jpv_16}, the emission spectrum  at $V=1$\,V based on the generalized Planck (GP) law \eqref{eq:genplanck} -- $\bar{\mathcal{R}}_{\textrm{em}}^{\textrm{GP}}$ -- does only coincide with the one obtained from the isotropic NEGF expression \eqref{eq:specemrate_av} -- $\bar{\mathcal{R}}_{\textrm{em}}^{\textrm{NEGF}}$ -- if the absorption coefficient at $V=1$\,V is used.  In contrast, the GP emission spectrum based on the absorption coefficient at short circuit conditions ($V=0$\,V) shows a strong field-induced red-shift and broadening as compared to the NEGF spectrum.

Since the validity of the GP law is an essential requisite for the photovoltaic reciprocity relation to hold, the above finding has severe consequences for the applicability of Expr.~\eqref{eq:recipro_norm} to the case of the ultrathin solar cells under consideration, as $Q_{\textrm{PV}}$ is conventionally defined at zero applied bias voltage where it provides the short circuit current $J_{sc}$ under the action of the external illumination spectrum. However, for a proper assessment of Expr.~\eqref{eq:recipro_norm}, the local relation \eqref{eq:genplanck} between absorption and emission first needs to be propagated to the \emph{global} relation between $Q_{\textrm{PV}}$ and the emitted photon flux $\phi_{\textrm{em}}$ at the surface of the device. To this end, in addition to the local dynamics, knowledge of the propagation of light inside the cell is required. This information is encoded in the NEGF version of the Poynting vector component $S_{z}(z)=\int dE_{\gamma}\,\mathcal{S}_{z}(z,E_{\gamma})$  for the energy flux normal to the slab surface,  which in terms of the photon GF is given by \cite{richter:08}
\begin{align}
\mathcal{S}_{z}(z,E_{\gamma})=&\frac{E_{\gamma}}{2\pi\hbar} \int \frac{d^{2}\mathbf{q}_{\parallel}}{(2\pi)^2} s_{z}(\mathbf{q}_{\parallel},z,E_{\gamma}),\\
s_{z}(\mathbf{q}_{\parallel},z,E_{\gamma})=&-\lim_{z'\rightarrow
z}\partial_{z'}\mathrm{Re}\sum_{\mu=x,y}\Big[\mathcal{D}_{\mu\mu}^{>}(\mathbf{q}_{\parallel},z,z',E_{\gamma})\nonumber\\&+
\mathcal{D}_{\mu\mu}^{<}(\mathbf{q}_{\parallel},z,z',E_{\gamma})\Big].\label{eq:modal_poynt}
\end{align}  
In Ref.~\onlinecite{richter:08}, Poynting's theorem for slab geometry $\partial_{z}S_{z}(z)=-W(z)$, with $W$ denoting the energy dissipation, was used to relate the (modal) absorptance of a homogeneous slab to the (modal) photon flux at the slab surface via
\begin{align}
S_{z}(z_{d})-S_{z}(z_{0})
=&-\int_{z_{0}}^{z_{d}}dz~W(z)\\
\equiv&-\int \frac{dE_{\gamma}}{2\pi\hbar} E_{\gamma}\int \frac{d^{2}\mathbf{q}_{\parallel}}{(2\pi)^2}w(\mathbf{q}_{\parallel},E_{\gamma})\label{eq:dissip}
\end{align}
where $z_{d}-z_{0}=d$ is the absorber thickness, and the modal dissipation can be written as  follows
\begin{align}
w(\mathbf{q}_{\parallel},E_{\gamma})=&-2\sum_{\mu,\nu}\big[b_{\mu\nu}(\mathbf{q}_{\parallel},E_{\gamma})-n_{\mu\nu}(\mathbf{q}_{\parallel},E_{\gamma})\big]\nonumber\\&\times a_{\mu\nu}(\mathbf{q}_{\parallel},E_{\gamma}).\label{eq:kirchhoff}
\end{align}
In the above expression, $b$ characterizes the global non-equilibrium distribution function of medium-induced fluctuations and $n$ is the distribution function of incident external photons \cite{richter:08}. The microscopic expression for the absorptance of the slab in the general non-equilibrium state is given by 
\begin{align}
a_{\mu\nu}(\mathbf{q}_{\parallel},E_{\gamma})=&-\int dz\int dz'~\Big[\hat{\mathcal{D}}_{v,\mu\nu}(\mathbf{q}_{\parallel},z,z',E_{\gamma})\nonumber\\
&\times\hat{\Pi}_{\nu\mu}(\mathbf{q}_{\parallel},z',z,E_{\gamma})\Big],\label{eq:absorpt_negf}
\end{align}
with $\hat{\mathcal{D}}_{v}$ the spectral function of vacuum-induced incident fluctuations~\cite{henneberger:09pra}. Under the assumption of complete isotropy, the absorptance obeys $a_{\mu\nu}(\mathbf{q}_{\parallel},E_{\gamma})\equiv \bar{a}(E_{\gamma})\delta_{\mu\nu}\cos\theta/2$ for $q_{\parallel}=q_{0}\sin\theta$  with $q_{0}=E_{\gamma}/(\hbar c_{0})$. In the case of global quasi-equilibrium characterized by a QFLS $\Delta \mu=\mu_{n}-\mu_{p}$, the distribution of medium-induced excitations amounts to the corresponding Bose-Einstein function, $b_{\mu\nu}(\mathbf{q}_{\parallel},E_{\gamma})\equiv f_{\textrm{BE}}(E_{\gamma}-\Delta\mu)$, and the integration in \eqref{eq:kirchhoff} is restricted to the absorptance. Hence, using Expr.~\eqref{eq:kirchhoff} for vanishing incident radiation field -- i.e., $n\equiv0$ -- Expr.~\eqref{eq:dissip} for emission into air becomes
\begin{align}
\phi_{\mathrm{em}}(E_{\gamma})=&~\mathcal{S}_{z}(z_{d},E_{\gamma})/E_{\gamma}=-\mathcal{S}_{z}(z_{0},E_{\gamma})/E_{\gamma}
 \\=&-\frac{1}{2} \int \frac{d^{2}\mathbf{q}_{\parallel}}{(2\pi)^3\hbar}w(\mathbf{q}_{\parallel},E_{\gamma})
 	\\=&~\bar{a}(E_{\gamma})\frac{E_{\gamma}^2}{4\pi^2\hbar^3c_{0}}f_{\textrm{BE}}(E_{\gamma}-\Delta\mu),
\end{align}
which is exactly the result found by W\"urfel in Ref.~\cite{wuerfel:82} for identical assumptions. 

\begin{figure}[t!]
	\begin{center}
		\includegraphics[width=0.5\textwidth]{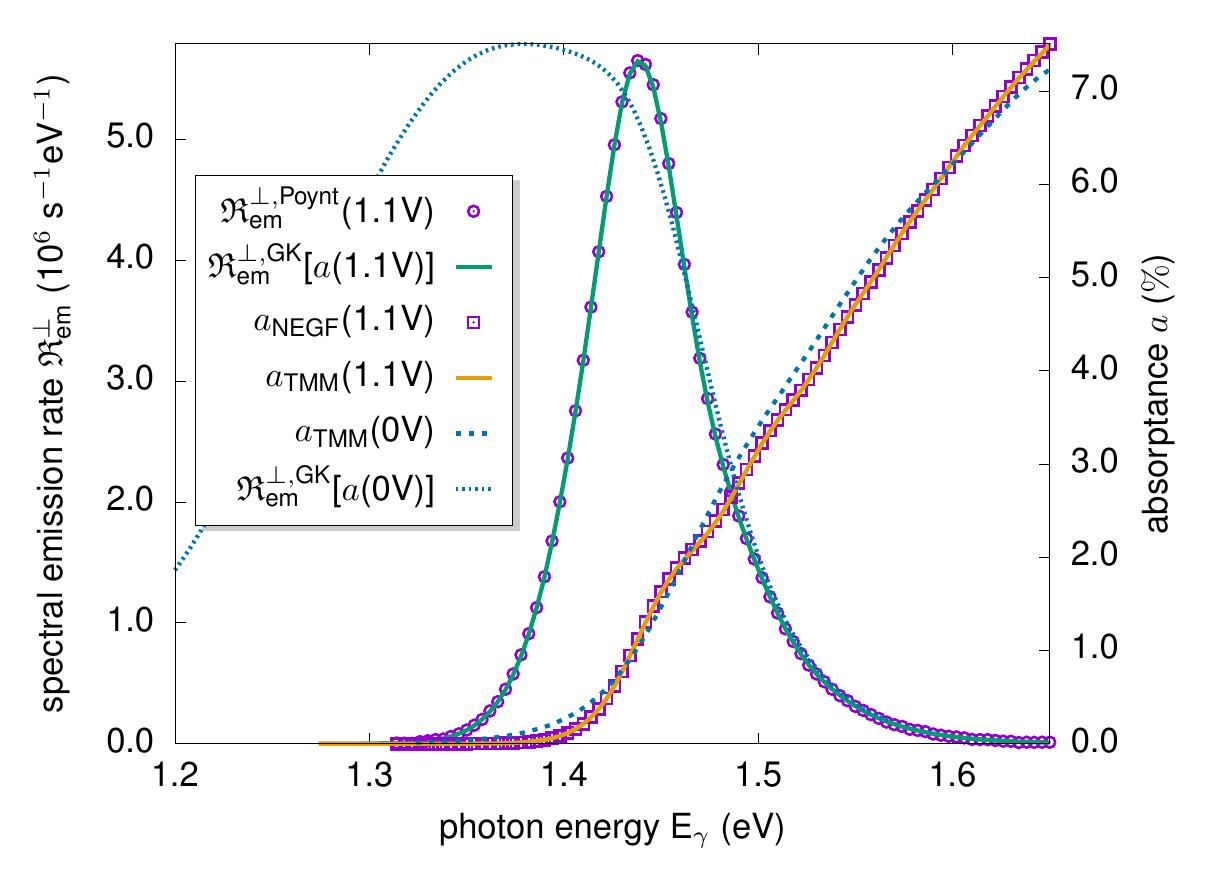}
		\caption{(color online) Total absorptance $a_{\mathrm{NEGF}}$ (open squares) and spectral emission rate $\mathfrak{R}_{\mathrm{em}}^{\perp,\mathrm{Poynt}}$ (open circles) of light with propagation direction normal to the left slab surface, for a bias voltage of $V=1.1$\,V applied at the contacts, as given by Eqs.~\eqref{eq:absorpt_negf} and \eqref{eq:emrate_poynt}.  The results coincide with the values obtained directly from the TMM and the generalized Kirchhoff (GK) law [Eq.~\eqref{eq:gk_perp}], if the actual absorptance of the biased system is used. For the absorptance at short circuit conditions ($V=0$\,V, dashed line), the GK emission spectrum (dotted line) deviates significantly from the actual NEGF emission spectrum, exhibiting again a strong red-shift and broadening attributed to the field-induced tailing of the joint density of states.   \label{fig:gk}}
	\end{center}
\end{figure}

\begin{figure}[t]
	\begin{center}
		\includegraphics[width=0.5\textwidth]{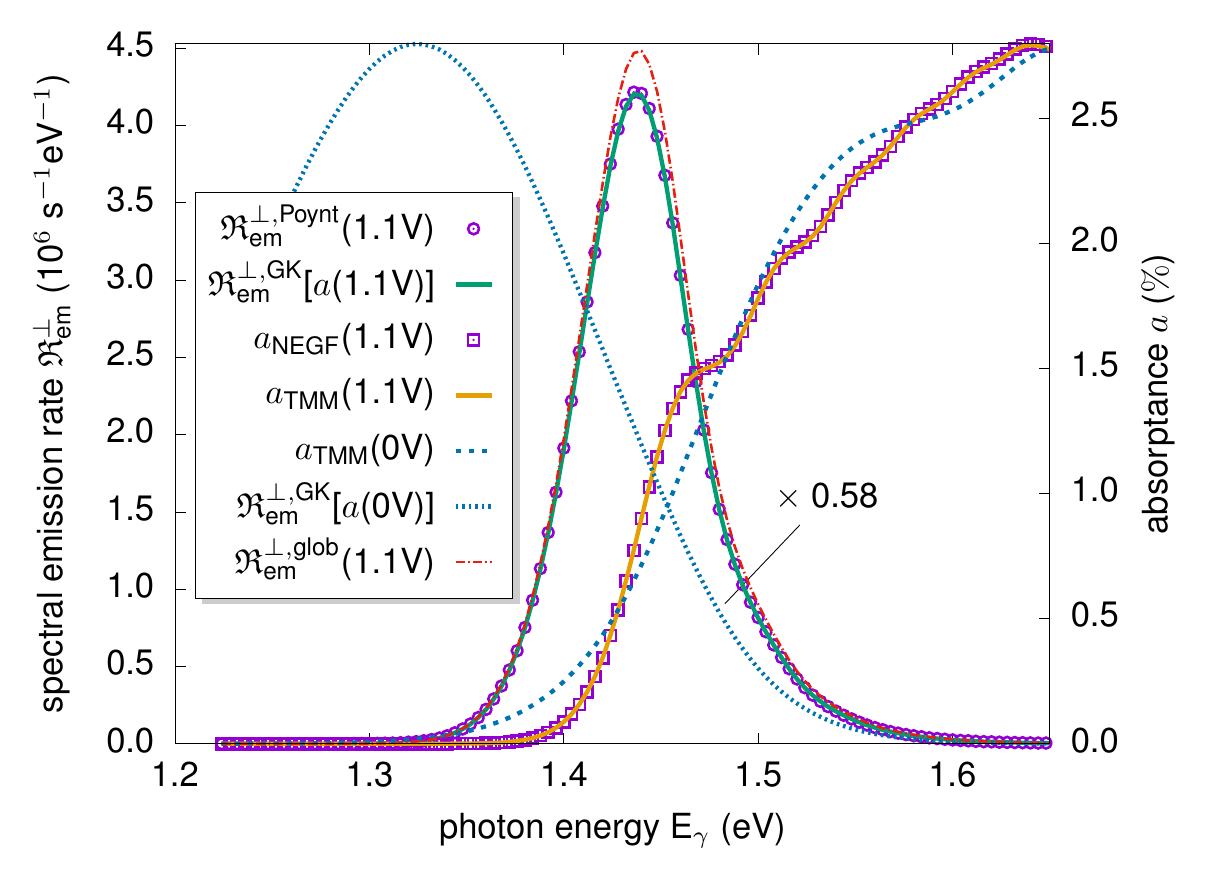}
		\caption{(color online) Same as Fig.~\ref{fig:gk}, but now for a system with a gold back reflector attached at the right side. Additionally, the global emission rate $\mathfrak{R}_{\textrm{em}}^{\perp,\textrm{glob}}$ obtained from the $z$-integration of the local rate for normal emission is shown (dash-dotted line), revealing the effect of reabsorption on the spectral shape and intensity of the emission at the surface.\label{fig:gk_refl}}
	\end{center}
\end{figure}

For the ultra-thin solar cell considered here, the global quasi-equilibrium conditions -- corresponding to practically constant QFLS throughout the absorber -- are well met, as can be inferred from Fig.~\ref{fig:bandprof}. The quasi-Fermi levels (QFL) $\mu_{n(p)}$ for electrons (holes) shown there are obtained from the NEGF carrier densities $\rho$ and spectral functions $\hat{G}\equiv i(G^{>}-G^{<})$ via the local fluctuation-dissipation theorem \cite{berbezier:15}, by solving numerically the non-linear equation 
\begin{align}
\rho_{n(p)}[G_{n(p)}^{\lessgtr}](z)=&\int \frac{dE}{2\pi}\int \frac{d^2\mathbf{k}_{\parallel}}{(2\pi)^2}\hat{G}_{n(p)}(\mathbf{k}_{\parallel},z,z,E)\nonumber\\\times&\{1/2\mp 1/2+f_{\textrm{FD}}[\mu_{n(p)}(z),E]\}
\end{align}
where $f_{\textrm{FD}}(\mu,E)=\{\exp[\beta(E-\mu)]+1\}^{-1}$ is the Fermi-Dirac distribution function at lattice temperature $T$ and the upper (lower) sign is for electrons (holes). The tiny gradient in the QFL reflects the high charge carrier mobility and the absence of fast recombination processes, and leads to the equivalence $Q_{\textrm{PV}}\equiv a$, as shown in Ref.~\onlinecite{ae:prb89_14}, where the short circuit current under monochromatic illumination is compared to the photocurrent obtained from the absorptance. The spectral rate of photon emission normal to the left surface into modes coupling to normally incident light as provided by 
\begin{align}
\mathfrak{R}_{\textrm{em}}^{\perp,\textrm{Poynt}}(E_{\gamma})=-s_{z}(\mathbf{0},z_{0},E_{\gamma})/(2\pi\hbar)\label{eq:emrate_poynt}
\end{align}
with the modal Poynting vector computed directly from the photon GF via \eqref{eq:modal_poynt} is therefore compared to the corresponding generalized  Kirchhoff law
\begin{align}
\mathfrak{R}_{\textrm{em}}^{\perp,\textrm{GK}}(E_{\gamma})=a(\mathbf{0},E_{\gamma})/(2\pi\hbar)f_{\textrm{BE}}(E_{\gamma}-\Delta\mu),\label{eq:gk_perp}
\end{align}
where the absorptance is given by \eqref{eq:absorpt_negf}. Figure \ref{fig:gk} displays the close agreement of $\mathfrak{R}_{\textrm{em}}^{\perp,\textrm{Poynt}}$ and  $\mathfrak{R}_{\textrm{em}}^{\perp,\textrm{GK}}$ for $\Delta\mu$ set to the applied bias voltage of $V=1.1$\,V. Also shown is the perfect match of the NEGF absorptance $a_{\textrm{NEGF}}$ used in $\mathfrak{R}_{\textrm{em}}^{\perp,\textrm{GK}}$ with the absorptance $a_{\textrm{TMM}}$ obtained from a transfer matrix method approach (TMM) using the absorption coefficient at $V=1.1$\,V, validating the photon component of the coupled NEGF approach. In analogy to the local relation between absorption coefficient and emission rate, the global emission spectra provided by the full NEGF solution are compared to those obtained from the generalized Kirchhoff (GK) law using the absorptance determined at short circuit conditions ($V=0$\,V), which corresponds to the standard definition of the external quantum efficiency $Q_{\textrm{PV}}$ for perfect carrier transport. Again, a large discrepancy in the form of a strong red-shift and broadening of $\mathfrak{R}_{\textrm{em}}^{\perp,\textrm{GK}}$[$\alpha$(0\,V)] as compared to $\mathfrak{R}_{\textrm{em}}^{\perp,\textrm{Poynt}}$ is observed, confirming the invalidity of Expr.~\eqref{eq:recipro_norm} in the regime of ultra-thin absorbers subject to large variation of built-in fields with applied bias voltage. This analysis holds also for the optically more complex situation of a device with a gold back reflector attached a the right side, as shown in Fig.~\ref{fig:gk_refl}. At this point, it is interesting to note that while the global emission rate resulting from spatial integration of the local rate \eqref{eq:locspecrate} gives the radiative dark current upon integration over photon energies, it does not coincide with the photon flux at the surface of the cell, as the propagation through the absorbing material is not accounted for. The resulting impact of reabsorption on the spectral shape and magnitude of the emission at the surface can be inferred from the difference of the curves for the emission rate given by the Poynting vector -- i.e., $\mathfrak{R}_{\textrm{em}}^{\perp,\textrm{Poynt}}$ -- and that obtained from the $z$-integrated modal emission rate \eqref{eq:locspecemmodrate}, i.e., $\mathfrak{R}_{\textrm{em}}^{\perp,\textrm{glob}}(E_{\gamma})=\int dz\,r_{\mathrm{em,spont}}(\mathbf{0},z,E_{\gamma})$ (dash-dotted line).

In conclusion, we presented a critical assessment of the photovoltaic reciprocity relation between external quantum efficiency and electroluminescent emission in ultrathin solar cells, by application of a comprehensive non-equilibrium quantum theory of photovoltaic device operation which considers both electronic and optical degrees of freedom on equal footing. The explicit relation of the macroscopic device properties to the microscopic nonequilibrium charge carrier states reveals the approximate nature of the semiclassical reciprocity theorem and provides at the same time a more generally valid picture of local (GP) and global (GK) connections between radiative processes in mesoscopic solar cell devices. This is of high practical relevance for photovoltaic devices whose characteristics are no longer determined by the equilibrium bulk properties of the constituent materials, but by the actual nonequilibrium device state at the operating point, as in the case of the ultrathin solar cell architectures considered here and in a wide range of nanostructure-based implementations of third-generation solar cell concepts.

\begin{acknowledgments}
 The authors acknowledge helpful discussions with Bart Pieters and Thomas Kirchartz. 
\end{acknowledgments}

\balance

%


\end{document}